%% file: ms.tex
\documentclass{llncs}

\usepackage{amsmath, amsfonts, graphicx, xcolor, url}
\usepackage{arydshln, multirow}
\usepackage{hyperref, footmisc, xurl}
\hypersetup{colorlinks=true, linkcolor=blue, urlcolor=blue, hyperfootnotes=true}

\newtheorem{customclaim}{Claim}

\usepackage[linesnumberedhidden,titlenotnumbered,ruled]{algorithm2e}
\SetKw{Continue}{continue}
\SetKw{Break}{break}
\SetKw{DownTo}{downto}
\newcommand{\MyComment}[1]{$/\ast$ #1 $\ast/$}

\newcommand\blfootnote[1]{%
	\begingroup
	\renewcommand\thefootnote{}\footnote{#1}%
	\addtocounter{footnote}{-1}%
	\endgroup
}

\begin{document}

\title{Cryptanalysis of the Privacy-Preserving Ride-Hailing Service TRACE}

\author{Deepak Kumaraswamy\inst{1}\orcidID{0000-1111-2222-3333} \and \\
	Srinivas Vivek \inst{2}\orcidID{0000-0002-8426-0859}}
\authorrunning{D. Kumaraswamy et al.}
% First names are abbreviated in the running head.
% If there are more than two authors, 'et al.' is used.
%
\institute{National Institute of Technology Karnataka, India \\
	\email{deepakkumaraswamy99@gmail.com}\\ \and
	International Institute of Information Technology Bangalore, India\\
	\email{srinivas.vivek@iiitb.ac.in}}

\maketitle

\blfootnote{© Springer Nature Switzerland AG 2021. The final published version is available at \url{www.springerlink.com}. DoI: 10.1007/978-3-030-92518-5}

\input{abstract}

\input{introduction}

\input{trace-overview}

\input{trace-attack}

\input{expt-results}

\input{related_work}

\input{conclusion}

\bibliographystyle{splncs04}
\bibliography{references}

\end{document}

%% file: abstract.tex
\begin{abstract}
In a typical ride-hailing service, the service provider (RS) matches a customer (RC) with the closest vehicle (RV) registered to this service. Ride-hailing services have gained tremendous popularity over the past years, and several works have been proposed to ensure privacy of riders and drivers during ride-matching. TRACE is an efficient privacy-preserving ride-hailing service proposed by Wang et al. (IEEE Trans.~Vehicular Technology 2018). TRACE uses masking along with other cryptographic techniques to ensure efficient and accurate ride-matching. RS computes a (secret) spatial division of a region into quadrants. The RS uses masked location information to match RCs and RVs within a quadrant without obtaining their exact locations, thus ensuring privacy. Additionally, an RC only gets to know location of the closest RV finally matched to it, and not of other responding RVs in the region. 

In this work, we disprove the privacy claims in TRACE by showing the following: a) RCs and RVs can identify the secret spatial division maintained by RS (this reveals information about the density of RVs in the region and other potential trade secrets), and b) the RS can identify exact locations of RCs and RVs (this violates location privacy). Prior to exchanging encrypted messages in the TRACE protocol, each entity masks the plaintext message with a secret unknown to others. Our attack allows other entities to recover this plaintext from the masked value by exploiting shared randomness used across different messages, that eventually leads to a system of linear equations in the unknown plaintexts. This holds even when all the participating entities are honest-but-curious. We implement our attack and demonstrate its efficiency and high success rate.
For the security parameters recommended for TRACE, an RV can recover the spatial division in less than a minute, and the RS can recover the location of an RV in less than a second on a commodity laptop.

\keywords{Location Privacy \and Privacy-Preserving Protocols \and Ride-Hailing Services \and Cryptanalysis \and Random Masking}

\end{abstract}

%% file: introduction.tex
\section{Introduction}
\label{sec:introduction}

Ride-hailing services such as Uber and Lyft have become a popular choice of transportation in the past decade \cite{pew:rhs_research}. By offering convenience and reliability to its customers, these services are well suited for intra-city commutes. A ride-hailing service usually consists of three entities: the ride-hailing server (RS), riders or customers (RCs) and drivers or vehicles (RVs). The RS is primarily responsible for hosting the ride-hailing service publicly. Drivers can register to this service and become identified as certified RVs. A customer who wishes to make use of this service can sign up as an RC and request for a ride. Depending on the pick-up and destination locations, the RS smartly forwards this ride request from RC to suitable RVs in the region. A list of nearby available RVs is revealed to the RC along with their reputations, who then makes a suitable choice.

However, revealing locations of RCs/RVs to other entities can have severe consequences. A pick-up location could correspond to the residential address of an RC, which can be used for stalking/kidnapping. There have also been instances when RVs registered to a particular ride-hailing service have been targeted by regular taxi drivers or targeted for theft \cite{thejournal,istanbul_uber}. Preserving privacy of sensitive users' locations has become a primary concern in ride-hailing services. 
Generally, the RS is assumed to be \textit{honest-but-curious}. This means that RS tries to learn as much information as possible without maliciously deviating from the ride-hailing protocol. 
Such a model is reasonable to assume since the RS wishes to preserve its reputation among the public. But it is still dangerous for the RS to learn locations of RCs and RVs, in case the RS later turns malicious or becomes a victim of cyberattacks \cite{et_datatheft,norton_databreach}.

In the past few years, there have been many works that focus on ensuring location privacy of RCs and RVs in the context of ride-hailing services. Section \ref{sec:related-work} contains an overview of recent papers in this area. These works use cryptographic primitives to hide sensitive location information from the RS, while trying to ensure efficiency and ride-matching accuracy. 

In this paper, we focus on TRACE \cite{tracewang2018}, proposed by Wang et.~al.~in 2018. TRACE is a privacy-preserving solution to ride-hailing services. Here, the RS first spatially divides each city into quadrants. RCs and RVs mask their sensitive location information using randomness and then forward it to RS. The RS then identifies the quadrant in which RCs and RVs lie, without finding out their exact locations. To ensure efficiency and accuracy, the ride request from an RC is forwarded only to RVs that are in the same quadrant as RC. The RC then makes a choice among RVs that lie in its vicinity to finalize ride establishment. Since the RS knows the distribution of RVs in different quadrants, it can periodically change its spatial division of the city to optimize bandwidth usage, reduce waiting time and improve accuracy.

TRACE uses masking with random secrets to prevent other entities of the protocol from learning the underlying message. At a high level, a large prime $p$ is chosen and the plaintext is multiplied with a random integer in $\mathbb{Z}_p$. These masked messages are encrypted using shared keys to prevent external eavesdroppers from gaining any useful information.
Since TRACE uses lightweight cryptographic techniques and simple modular arithmetic, it is efficient in practice. The security guarantees for TRACE state that RS cannot learn about the exact locations of RCs and RVs apart from the quadrant they are in. Additionally, RCs and RVs cannot learn about the secret spatial division maintained by RS, since this could reveal the density of drivers across the city, among other proprietary information and trade secrets of RS.

\subsection{Our Contribution}
\label{sec:our-contribution}

We propose an attack on TRACE and disprove the above security claims by showing that the RS can indeed retrieve the exact locations of all RCs and RVs. Secondly, we show that RCs and RVs can learn the secret spatial division information maintained by RS. These attacks constitute a total break of the privacy objectives of TRACE.
The underlying idea behind our attack is to eliminate the (unknown) randomness shared across different messages when other entities mask their location values. This allows one to efficiently obtain an overdetermined system of linear (modular) equations in the unknown plaintext locations.
We stress that this attack is purely algebraic, and does not make any geometric assumptions about the region.
Our attack is efficient (runs in time quadratic in the security parameters) and holds even when all entities are honest-but-curious. 
For instance, with the recommended security parameters from \cite{tracewang2018}, an RV can recover the quadtree maintained by RS in under a minute (see Table \ref{tab:rs-quadtree-timing}) and the RS can recover the exact location of an RV in under a second (see Table \ref{tab:rs-rv-timing}).

The rest of our paper is organized as follows. In Section \ref{sec:trace-overview}, we describe relevant steps of the TRACE protocol from \cite{tracewang2018}. 
%This also serves as an overview for new readers who wish to understand TRACE. 
The first attack in Section \ref{sec:attack-rc-rv-quadtree} describes how RCs and RVs can recover the secret quadtree maintained by RS. The second attack in Section \ref{sec:attack-rs-location} describes how the RS can recover exact locations of RCs and RVs. We briefly discuss a modification to the TRACE protocol that prevents only the first attack, and argue that the second attack (which is more severe than the first) is hard to thwart.
Algorithms \ref{algo:rv-quadtree} and \ref{algo:rs-rv-location} summarize the above two attacks.
Section \ref{sec:experimental-results} provides details about our experimental setup and evaluates the efficiency and success rate of our attack in practice (refer Tables \ref{tab:rs-quadtree-timing} and \ref{tab:rs-rv-timing}). Section \ref{sec:related-work} gives an overview of recent works in the area of privacy-preserving ride-hailing services. We conclude our paper and provide remarks about future work in Section \ref{sec:conclusion-future-work}.

%% file: trace-overview.tex
\section{Overview of TRACE}
\label{sec:trace-overview}

This section contains a high level overview of the TRACE protocol \cite{tracewang2018}. Details that are not directly relevant to our attack will be omitted. For more information the reader is referred to the original paper.

\subsection{Preliminaries}
\label{sec:trace-preliminaries}

\begin{figure}[h]
    \centering
    \includegraphics[width=3.3in]{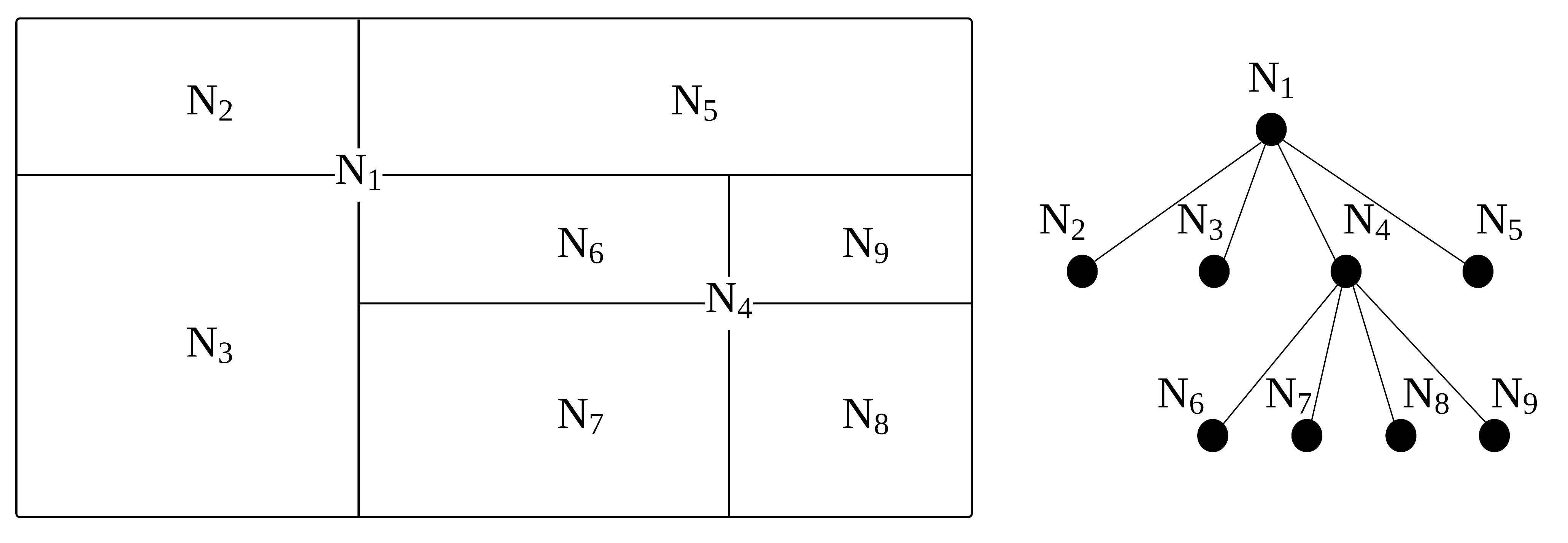}
    \caption{Example of a quadtree}
    \label{fig:quadtree}
\end{figure}

A quadtree $\{N_1,\dots,N_m\}$ with $m$ nodes is a data structure used to represent the partition of a 2-D space into quadrants and subquadrants. Each node $N_i$ in the tree is associated with four $(x,y)$ coordinates denoting corners of the quadrant represented by that node. Every non-leaf node in the quadtree has four children denoting the division of that quadrant into four subquadrants. 
An example is presented in Figure \ref{fig:quadtree}.

Given a point $P=(x,y)$ and a quadrant $\{(x_j,y_j)\}$ with $j=1,\dots ,4$, we can easily check if $P$ lies within the quadrant by doing the following \cite[Section \uppercase\expandafter{\romannumeral 3}]{tracewang2018}. For each $j$ compute 
\begin{equation}
\label{eq:lie-quad}
S_j = (xy_j + yx_{j'} + x_jy_{j'}) - (xy_{j'} + yx_j + x_{j'}y_j)
\end{equation} 
where $j'=(j \mod 4)+1$. If all $S_j \geq 0$, then $P$ lies within the quadrant, otherwise it does not. Given a quadtree, this idea can be extended to find the quadrant/node of the tree in which $P$ lies. Starting at the root, among its four children, find that quadrant/node in which $P$ lies; then recurse on its children until a leaf is encountered.

\subsection{System Design and Security Goals}
\label{sec:trace-security-model}

\textbf{System Design.}
The three primary entities in the TRACE protocol are the ride-hailing server/service provider (RS), the customer/rider (RC) and the vehicle/driver (RV). All of the aforementioned entities are assumed to be honest-but-curious. This means that they wish to learn as much information as they can about the other entities without violating any protocol steps. 

RS is mainly responsible for forwarding requests/responses between RCs and RVs. As part of the protocol, RS maintains a spatial division of the city into quadrants and uses it to identify regions in which RCs and RVs lie. It does so in such a way that RCs and RVs do not learn any information about the spatial division, while RS does not learn the exact locations of RCs and RVs. The RC can choose a pick-up point and send a ride-hailing request to RS, who then forwards it to the RVs that lie in close vicinity of RC. RVs submit their masked location information to RS at regular intervals, allowing the RS to have an idea of distribution of RVs in the city. Depending on the density of RVs, RS can periodically optimize its space division to improve ride-matching accuracy. 

\textbf{Threat Model.}
We assume the same threat model that is considered in TRACE. All entities are assumed to be honest-but-curious, that is, they follow the protocol specification but may infer additional data from the observed transcripts. RS does not collude with RCs and RVs (to try and obtain information about customers), since it has an incentive to maintain high reputation.

\textbf{Security Goals.}
It is essential to ensure that location information of RCs and RVs is not revealed to other entities. The spatial division maintained by RS should also be kept secret, as this could reveal information about density of drivers in a city and other proprietary information/trade secrets of RS. The authors of TRACE claim that the following security requirements are satisfied during the protocol execution.

\begin{customclaim}
	\label{claim:rs-spatial}
	RS creates a quadtree $N$ containing information about spatial divison of the city into quadrants, and masks it with a randomly chosen secret to compute $EN$. Given $EN$, RCs and RVs do not learn anything about $N$.
\end{customclaim}

\begin{customclaim}
	\label{claim:rs-rc-location}
	RS can only learn the quadrants in which RCs lie. RS does not obtain any other information about the exact pick-up locations of RCs.
\end{customclaim}
    
\begin{customclaim}
\label{claim:rs-rv-location}
RS can only learn the quadrants in which RVs lie. RS does not obtain any other information about the exact locations of RVs.
\end{customclaim}

%\begin{customclaim}
%   \label{claim:rc-rv-location}
%   At the end of ride-matching, RC gets to know the location of only that RV which is matched to it. RC does not learn about the exact locations of other RVs that are in its vicinity.
%\end{customclaim}

\subsection{TRACE Protocol}
\label{sec:trace-protocol}

\begin{figure}
	\centering
	\includegraphics[width=3in]{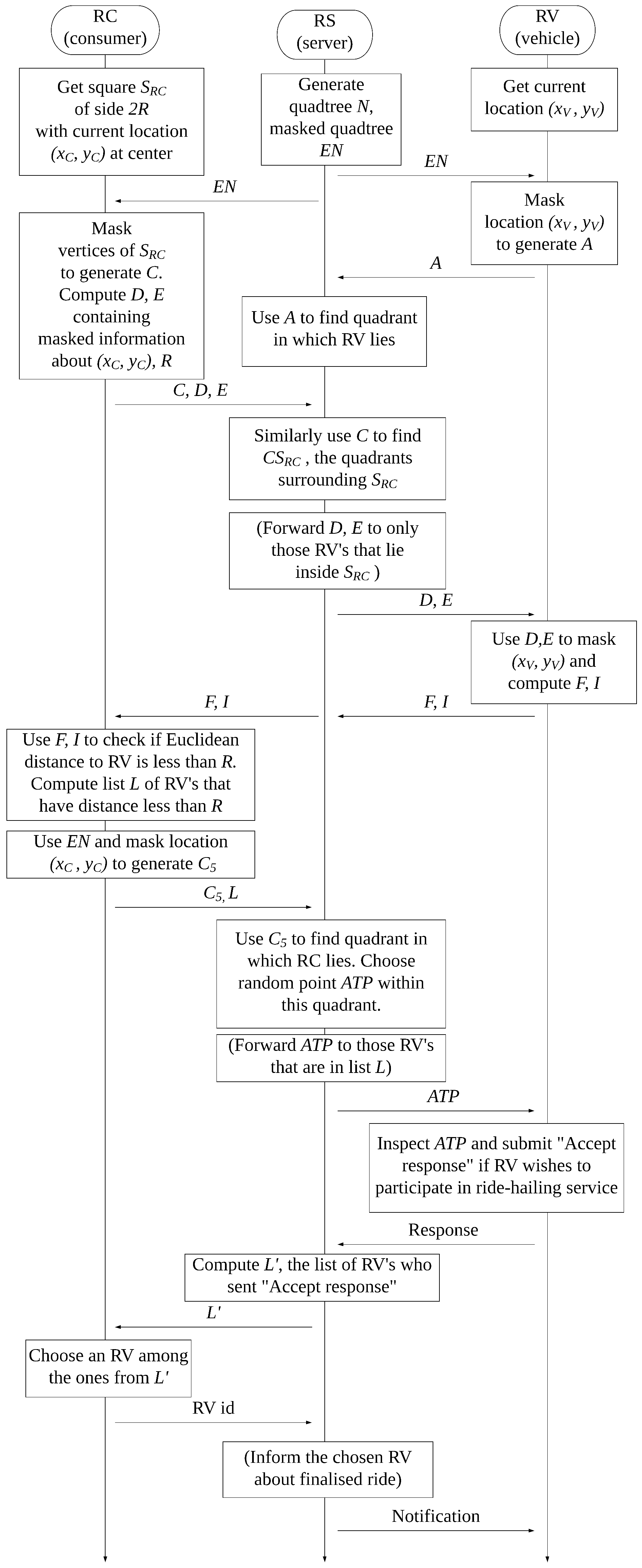}
	\caption{Overview of TRACE protocol}
	\label{fig:trace-protocol}
\end{figure}

This section describes the execution of the TRACE protocol. Figure \ref{fig:trace-protocol} gives a summarized view of the messages exchanged between different entities. 
RS acts as a central entity for forwarding messages between RCs and RVs. It establishes shared keys with RCs and RVs through the Diffie-Hellman key exchange. All messages exchanged between RS and RCs, RVs are encrypted using a symmetric encryption scheme. The authentication of entities is ensured by signing these messages using the BLS signature scheme \cite{blssignature2001}. The notations used in the TRACE protocol and their descriptions are provided in Table \ref{tab:trace-notations}.

\begin{table}[h]
	\renewcommand{\arraystretch}{1.5}
	\centering
	\begin{tabular}{|c|c|}
		\hline
		\textbf{Notation} & \textbf{Description}   \\
		\hline
		RS & Ride-hailing server (service provider) \\
		\hline RC & Customer (rider) \\
		\hline RV & Vehicle (driver) \\
		\hline $k_1, k_2, k_3, k_4$ & Security parameters of TRACE \\
		\hline $N$ & Spatial division (quadtree) maintained by RS \\
		\hline $\alpha, p$ & Large primes chosen by RS \\
		\hline $\alpha', p'$ & Large primes chosen by RC \\
		\hline $(x_{Nij},y_{Nij})$ & Coordinates of $j$-th vertex in the $i$-th quadtree node $N_i$ \\
		\hline $s, a_{jh}$ & Random values used by RS when masking $N$ \\
		\hline $EN$ & Masked quadtree computed by RS \\
		\hline $(x_V, y_V)$ & Coordinates of RV \\
		\hline $r_{ij}$ & Random values chosen by RV when masking $(x_V, y_V)$ \\
		\hline $\pi(\cdot) $ & Random permutation chosen by RV \\
		\hline $A$ & Data aggregated by masking $(x_V,y_V)$ and $EN$ \\
		\hline $(x_{CP}, y_{CP})$ & Pick-up coordinates of RC \\
		\hline $S_{RC}$ & Square with $(x_{CP}, y_{CP})$ at its center \\
		\hline $R$ & Length of a side of $S_{RC}$ \\
		\hline $s', d_i$ & Random values used by RC when masking $(x_{CP}, y_{CP})$ \\
		\hline $C_1,\dots,C_5$ & Data aggregated by masking $(x_{CP},y_{CP})$ and $EN$ \\
		\hline
	\end{tabular}
	\vspace{0.1in}
	\caption{Description of the notations used in the TRACE protocol.}
	\label{tab:trace-notations}
\end{table}

For convenience, the remainder of this paper shall refer to subscripts $ (\cdot)_{i,j} $ and $ (\cdot)_{i,j,l} $ as simply $ (\cdot)_{ij} $ and $ (\cdot)_{ijl} $, respectively.

\textbf{Step 0.} RS publishes details about different system parameters (for example, the group and its generator used in the signature scheme, public key of RS, choice of symmetric encryption). RCs and RVs also establish their public keys. RS announces security parameters $k_1,k_2,k_3,k_4$. As we shall see subsequently, they specify the size of different randomness used when masking location information. Step 3 elaborates on the constraint that should exist among these four parameters to ensure correctness of the protocol.

RS chooses two large public primes $p$ and $\alpha$ (of size $k_1$ bits and $k_2$ bits, respectively) and a random secret $s\in \mathbb{Z}^*_p$ known only to itself.

\textbf{Step 1.} RS divides the two-dimensional space into squares or rectangles represented by a quadtree \[
N=\{N_1,N_2,\dots,N_m\}
\] 
with $m$ nodes. The $i$-th quadrant $N_i$ has four corners $\{N_{ij}=(x_{Nij},\allowbreak y_{Nij})\}$ where $j=1,\dots ,4$.
RS wishes to learn the quadrant in which each RV lies without learning its exact location. To do this, RS sends a masked version of $N_i$ to RV. Concretely, RS chooses 24 random values $a_{jh}$ ($j=1,\dots ,4;h=1,\dots ,6$) of size $k_3$ bits each. For every vertex
% $(x_{Nij}, y_{Nij})$
$N_{ij}$ of $N_i$, let 
%$(x_{Nij'}, y_{Nij'})$
$N_{ij'}$ be the vertex adjacent to it in the anticlockwise direction, i.e.~$j'=(j\mod 4)+1$. RS masks this vertex by computing
\begin{align*}
EN_{ij1} &= s(x_{Nij} \cdot \alpha + a_{j1})\mod p~, \\
EN_{ij2} &= s(y_{Nij} \cdot \alpha + a_{j2})\mod p~, \\
EN_{ij3} &= s(x_{Nij'} \cdot \alpha + a_{j3})\mod p~, \\
EN_{ij4} &= s(y_{Nij'} \cdot \alpha + a_{j4})\mod p~, \\
EN_{ij5} &= s(x_{Nij} \cdot y_{Nij'} \cdot \alpha + a_{j5})\mod p~, \\
EN_{ij6} &= s(x_{Nij'} \cdot y_{Nij} \cdot \alpha + a_{j6})\mod p~.
\end{align*}
The values $\alpha, p$ are public, whereas $s,x_{Nij},y_{Nij}$ are only known to RS. The masked coordinate is 
\[
EN_{ij}=EN_{ij1} \allowbreak \| \allowbreak EN_{ij2} \| \allowbreak EN_{ij3} \| \allowbreak EN_{ij4} \| \allowbreak EN_{ij5} \| \allowbreak EN_{ij6}~,
\] 
where $\|$ denotes concatenation. Next, RS computes the masked quadrant  
\[
EN_i=EN_{i1}\| \allowbreak EN_{i2}\| \allowbreak EN_{i3}\| \allowbreak EN_{i4},
\] 
for $i=1,\dots,m$, to get the masked quadtree 
\[ 
EN=\{EN_1,EN_2,\dots,EN_m\}.
\]
It then encrypts $EN$ and forwards it to RV.

\textbf{Step 2.} RV decrypts this message and uses $EN$ along with its own randomness to mask its location $(x_V,y_V)$. For $i=1,\dots ,m;j=1\dots,4$, RV chooses a fresh random number $r_{ij}$ (each $k_4$ bits long) and computes
\begin{align*}
A_{ij1} &= r_{ij} \cdot \alpha (x_V\cdot EN_{ij4} + y_V \cdot EN_{ij1} + EN_{ij6}) \mod p~, \\
A_{ij2} &= r_{ij} \cdot \alpha (x_V\cdot EN_{ij2} + y_V \cdot EN_{ij3} + EN_{ij5}) \mod p~, \\
A_{ij} &= A_{ij1}\| A_{ij2}~.
\end{align*}

RV chooses a random permutation $\pi(\cdot)$ to reorder the $j$-indices for each $A_i$. That is, 
\begin{align*}
A_i & =  A_{i\pi(1)} \| A_{i\pi(2)} \| A_{i\pi(3)} \| A_{i\pi(4)} , \\ 
A & =  \{A_1,\dots,A_m\}.
\end{align*}
The order within each $A_{i\pi(j)}$ is still preserved, that is, 
\[ 
A_{i\pi(j)}=A_{i\pi(j)1} \| A_{i\pi(j)2}.
\]
RV encrypts $A$ and forwards it to RS. 

\textbf{Step 3.} RS obtains $A$ that contains the masked location of each RV, and does the following computations to identify the quadrant/node $N_i$ of the quadtree in which RV lies.

\begin{align*}
B_{ij1} &= s^{-1} \cdot A_{ij1} \mod p  \\
&= s^{-1} \cdot r_{ij} \cdot \alpha (x_V\cdot EN_{ij4} + y_V \cdot EN_{ij1} + EN_{ij6}) \mod p \\
&= s^{-1} \cdot r_{ij} \cdot s [{\alpha}^2(x_V \cdot y_{Nij'} + y_V \cdot x_{Nij} + x_{Nij'} \cdot y_{Nij}) \\
& + \alpha(x_V \cdot a_{j4} + y_V \cdot a_{j1} + a_{j6})] \mod p~. \\
B'_{ij1} &= \frac{B_{ij1} - (B_{ij1} \mod \alpha^2)}{\alpha^2} \\
&= r_{ij}(x_V \cdot y_{Nij'} + y_V \cdot x_{Nij} + x_{Nij'} \cdot y_{Nij})~. \\
\end{align*}
Similarly,
\begin{align*}
B_{ij2} &= s^{-1} \cdot A_{ij2} \mod p~,  \\
B'_{ij2} &= \frac{B_{ij2} - (B_{ij2} \mod \alpha^2)}{\alpha^2} \\
&= r_{ij}(x_V \cdot y_{Nij} + y_V \cdot x_{Nij'} + x_{Nij} \cdot y_{Nij'})~.
\end{align*}
Next, RS computes the difference
\begin{align*}
B_{ij} &= B'_{ij2} - B'_{ij1} \\
&= r_{ij} [(x_V \cdot y_{Nij} + y_V \cdot x_{Nij'} + x_{Nij} \cdot y_{Nij'}) \\
&- (x_V \cdot y_{Nij'} + y_V \cdot x_{Nij} + x_{Nij'} \cdot y_{Nij})]~.
\end{align*}

Compare this to Equation \eqref{eq:lie-quad}. 
Since $r_{ij}$ is always positive, RS can identify whether RV lies in $N_i$ by checking if $B_{ij}$ is positive for all $j=1,\dots, 4$. Using the method described in Section \ref{sec:trace-preliminaries}, RS can query the quadtree to identify the exact quadrant where RV lies.

Note that it was necessary to remove the modulus with respect to $p$ when obtaining $B'_{ij1}$ and $B'_{ij2}$, otherwise those values would always be positive irrespective of whether RV was inside the quadrant $N_i$ or not. To remove this modulus it is sufficient if the following is always true during the computation of $B_{ij1}$ (a similar condition exists for $B_{ij2}$).
\begin{align*}
&r_{ij}[{\alpha}^2(x_V \cdot y_{Nij'} + y_V \cdot x_{Nij} + x_{Nij'} \cdot y_{Nij}) \\
& + \alpha(x_V \cdot a_{j4} + y_V \cdot a_{j1} + a_{j6})] < p~,\\
& r_{ij}\cdot \alpha (x_V \cdot a_{j4} + y_V \cdot a_{j1} + a_{j6}) < \alpha^2~,\\
&r_{ij}\cdot \alpha (x_V \cdot a_{j2} + y_V \cdot a_{j3} + a_{j5}) < \alpha^2~.
\end{align*}
Let $\langle \cdot \rangle$ denote the bit length of a non-negative integer. Recall that $\langle p \rangle=k_1,\allowbreak \langle \alpha \rangle=k_2,\allowbreak \langle a_{jh} \rangle=k_3,\allowbreak \langle r_{ij} \rangle=k_4$. To ensure the above conditions hold, the parameters are chosen such that
\begin{align}
k_4 + 2k_2 < k_1~,  \nonumber \\ 
k_2 + k_3 < k_1~, \nonumber \\ 
k_3 + k_4 < k_2~. \label{eq:security-params}
\end{align}
Moreover, the size of location coordinates are assumed to be negligible compared to these security parameters. In \cite{tracewang2018}, the above values are set as $k_1=512,k_2=160,k_3=75,k_4=75$. 

\textbf{Step 4.} RC receives $EN$ from RS. Now the RC tries to mask its location with respect to the quadtree and send it to RS. Suppose the pick-up point of RC is $ (x_{CP}, y_{CP}) $. RC chooses a square $S_{RC}$ of side $2R$ (where $R$ is $\geq$ 1 km) with this pickup point at its center. Let the vertices of this square be $ \{ (x_{S1}, y_{S1}), \allowbreak (x_{S2}, y_{S2}), \allowbreak (x_{S3}, y_{S3}), \allowbreak (x_{S4}, y_{S4}) \} $. Recall that in Step 2, each RV masked its location $(x_V, y_V)$ with respect to $EN$ and computed $A$. RC also does an equivalent computation here; after receiving $EN$ from RS, it computes a masking for each of the four vertices of $S_{RC}$ to obtain $C=C_1\|C_2\|C_3\|C_4$.

Next, RC chooses a public prime $p'$ of size $k_1$ bits, a public prime $\alpha'$ of $k_2$ bits, a secret $s' \in \mathbb{Z}^*_{p'}$ and 4 random values $d_i$ of $k_4$ bits each. It computes 
\begin{align*}
D_1 &= s'(x_{CP}\cdot \alpha' + d_1) \mod p'~, \\
D_2 &= s'(y_{CP}\cdot \alpha' + d_2) \mod p'~, \\
D_3 &= s'\cdot d_3 \mod p'~,\\
D_4 &= s'\cdot d_4 \mod p'~,\\
D &= D_1 \| D_2\|D_3\|D_4~,\\
E &= x_{CP}^2 + y_{CP}^2 - R^2~.
\end{align*}
RC encrypts $C,D,E$ and sends it to RS.

\textbf{Step 5.} The goal here is to convey the masked location information from RC to RVs that are ``nearby'' to it. RS decrypts the message from RC to get $C,D,E$. Similar to Step 3, for each of $C_1,C_2,C_3,C_4$, RS obtains the quadrant in which the vertex represented by $C_i$ (i.e.~ $ (x_{Si},y_{Si}) $) lies. With this RS knows the quadrants in which the corners of square $S_{RC}$ lies. RS can construct a region $CS_{RC}$ enclosing $S_{RC}$. From Step 3, RS also knows the quadrants in which each RV lies. RS encrypts $D,E$ and sends it to those RVs that lie in $CS_{RC}$ (call these RVs as SRVs).

\textbf{Step 6.} SRV receives $D,E$ from RS and tries to add in masked information about its own location $(x_{SV}, y_{SV})$ to these values. It chooses three random $r_i$'s of $k_4$ bits each and computes
\begin{align*}
F_1 &= x_{SV} \cdot \alpha' \cdot D_1 \mod p'~, \\
F_2 &= y_{SV} \cdot \alpha' \cdot D_2 \mod p'~, \\
F_3 &= r_1 \cdot D_3 \mod p'~,\\
F_4 &= r_2 \cdot D_4 \mod p'~,\\
F &= r_3 (F_1+F_2+F_3+F_4)~,\\
I &= r_3 (x_{SV}^2 + y_{SV}^2 + E)~.
\end{align*}
SRV encrypts and sends $I,F$ to RC via RS.

\textbf{Step 7.} RC uses $I,F$ (that contain masked information of RC's and SRVs' locations) to check if that SRV is within distance $R$.
\begin{align*}
J &= {s'}^{-1}\cdot F \mod p'\\
&= {s'}^{-1}\cdot s' \cdot r_3 [{\alpha'}^2(x_{CP}\cdot x_{SV} + y_{CP}\cdot y_{SV}) \\
&\quad+ \alpha' (x_{SV}\cdot d_1 + y_{SV}\cdot d_2) + r_1\cdot d_3 + r_2\cdot d_4] \mod p'~,\\
J' &= \frac{J - (J\mod {\alpha'}^2)}{{\alpha'}^2} = r_3(x_{CP}\cdot x_{SV} + y_{CP}\cdot y_{SV})~, \\
K &= I - 2J'\\
&= r_3[x_{CP}^2 + y_{CP}^2 + x_{SV}^2 + y_{SV}^2  \\
&\quad - 2(x_{CP}\cdot x_{SV} + y_{CP}\cdot y_{SV}) - R^2]\\
&= r_3[(x_{CP}-x_{SV})^2 - (y_{CP}-y_{SV})^2 - R^2]~.
\end{align*}
When $K\leq 0$, the SRV is within the circle query range $C_{RC}$ of radius $R$ around $RC$. Call such SRVs as CRVs.

Once again (similar to Step 3) we need to eliminate the modulus with respect to $p'$ (otherwise $K$ would always be positive even if the SRV had distance $>R$). With the relationship imposed on the security parameters (Equation \eqref{eq:security-params} in Step 3), the following condition holds and the modulus is removed.
\begin{align*}
&r_3 [{\alpha'}^2(x_{CP}\cdot x_{SV} + y_{CP}\cdot y_{SV}) \\
&+ \alpha' (x_{SV}\cdot d_1 + y_{SV}\cdot d_2) + r_1\cdot d_3 + r_2\cdot d_4] < p'~, \\
&r_5[\alpha' (x_{SV}\cdot d_1 + y_{SV}\cdot d_2) + r_1\cdot d_3 + r_2\cdot d_4] < {\alpha'}^2~.
\end{align*}

\textbf{Step 8.} RC masks its take-off point $(x_{CT}, y_{CT})$ using $EN$ (similar to Step 2) to create $C_5$, and forwards $C_5$ along with the list of CRVs to RS. (The take-off point usually lies very close to the RC's pick-up point from Step 4). Similar to Step 3, RS uses $C_5$ to identify the subregion in which the take-off point lies. RS chooses a random location $ATP$ in this subregion and forwards it to CRVs. Each CRV inspects $ATP$ to make a decision on whether to accept this ride-hailing request from RC. The CRVs who decide to accept send an ``Accept Response'' to RS. RS forwards the list of ready and available CRVs to RC. RC chooses a suitable CRV from this list, and this CRV is informed about the same by RS. Later, the RC and the chosen CRV proceed with ride establishment by negotiating a shared session key and by exchanging information such as location, phone number, reputation, etc.

%% file: trace-attack.tex
\section{Attack on TRACE}
\label{sec:attack-trace}

This section presents two attacks which (with high empirical probability) disprove the following privacy claims made about TRACE.
First, in Section \ref{sec:attack-rc-rv-quadtree}, we show that RCs and RVs can obtain the secret spatial division (quadtree) information maintained by RS (violation of Claim \ref{claim:rs-spatial}). We also discuss a modification to the TRACE protocol, as a countermeasure for this attack. 
Secondly, in Section \ref{sec:attack-rs-location},  we show how the RS can identify exact locations of all RCs and RVs (violation of Claims \ref{claim:rs-rc-location}, \ref{claim:rs-rv-location}). We also briefly argue why this attack is not straightforward to thwart. In both attacks, the entities recover location coordinates modulo prime $p$. This is same as recovering the actual integer values since $p$ is a very large prime and the coordinate values are negligibly small compared to $p$.

Steps from the TRACE protocol described in Section \ref{sec:trace-protocol} will be referred as and when needed.
In Section \ref{sec:experimental-results}, we shall experimentally evaluate the success probability of our attacks.

\subsection{RCs, RVs obtain Quadtree}
\label{sec:attack-rc-rv-quadtree}

After an RV receives the masked quadtree $EN$ computed by RS (Step 2), we show how it can recover all underlying vertices $x_{Nij},y_{Nij}$ of the quadtree's nodes. This same principle allows an RC to obtain information about the quadtree as well (recall that each RC receives $EN$ from RS in Step 4).

\textbf{Intuition.} Intuitively, our attack works as follows. Each quadtree node $N_i$ is masked by the RS using random values $s, \alpha, a_{jh}$, resulting in $EN_i$. When an RV receives $EN_1,\dots,\allowbreak EN_m$, it knows $p,\alpha$ but does not know $s,a_{jh}$. For a single $EN_i$, the number of equations involved is $4\times 6=24$ (since there is one equation for each $EN_{ijh}$, $j=1,\dots,4; h=1,\dots,6$). The number of unknowns involved in $EN_i$ is $1+24+8=33$ ($s$, $a_{jh}$'s and quadrant vertices $x_{Nij},y_{Nij}$, $j=1,\dots,4; h=1,\dots,6$). A key observation is that if one considers $EN_i$ along with a different $EN_{i'}$, the number of equations is $24+24=48$. However the number of unknowns involved is $1+ 24 + 8 + 8 = 41$ ($s$, $a_{jh}$'s and quadrant vertices $x_{Nij},y_{Nij},x_{Ni'j},y_{Ni'j}$, where $j=1,\dots,4; h=1,\dots,6$). That is, considering an additional $EN_{i'}$ gives 24 new equations but introduces only 8 new variables.
This would allow RV to solve this system of modular equations and obtain the secrets $s$ along with quadrant vertices of $N_i$ and $N_{i'}$. 

\textbf{Formal attack.} Without loss of generality, we show how an RV can recover vertices of quadrants $N_1,N_2$ when given $EN_{1},EN_{2}$ (i.e.~$i=1,i'=2$). The first task is to eliminate the unknown randomness $a_{jh}, \enspace j=1,\dots, 4; h=1,\dots,6$. This can be done by subtracting $EN_{2jh}$ from $EN_{1jh}$.
For $h=1,\dots, 6$, we get the following equations.
\begin{align}
EN_{1j1} - EN_{2j1} &= s\alpha(x_{N1j} - x_{N2j}) \mod p~, \label{eq:en1} \\
EN_{1j2} - EN_{2j2} &= s\alpha(y_{N1j} - y_{N2j}) \mod p~, \label{eq:en2} \\
EN_{1j3} - EN_{2j3} &= s\alpha(x_{N1j'} - x_{N2j'}) \mod p~, \label{eq:en3} \\
EN_{1j4} - EN_{2j4} &= s\alpha(y_{N1j'} - y_{N2j'}) \mod p~, \label{eq:en4} \\
EN_{1j5} - EN_{2j5} &= s\alpha(x_{N1j}y_{N1j'} - x_{N2j}y_{N2j'}) \mod p~, \label{eq:en5} \\
EN_{1j6} - EN_{2j6} &= s\alpha(x_{N1j'}y_{N1j} - x_{N2j'}y_{N2j}) \mod p~. \label{eq:en6} 
\end{align}
Here $j'=(j\mod 4)+1$. The parameters $s,\alpha$ are unknown to RV along with the 16 variables $x_{N1j},y_{N1j},x_{N2j},y_{N2j}, \enspace j=1,\dots, 4$. RV can obtain linear (modular) equations in these variables by eliminating $s,\alpha$ as follows.

Compare $\eqref{eq:en1}\times y_{N1j'} + \eqref{eq:en4}\times x_{N2j}$ and $\eqref{eq:en5}$: 
\begin{align}
&(EN_{1j1} - EN_{2j1}) \times y_{N1j'} + (EN_{1j4} - EN_{2j4})\times x_{N2j} \nonumber \\
&= s\alpha(x_{N1j}y_{N1j'} - x_{N2j}y_{N1j'} + y_{N1j'}x_{N2j} - y_{N2j'}x_{N2j}) \nonumber \\
&= s\alpha(x_{N1j}y_{N1j'} - x_{N2j}y_{N2j'}) \nonumber \\
&= (EN_{1j5} - EN_{2j5}) \mod p \label{eq:lin-xy-1}~.
\end{align}

Compare $\eqref{eq:en1}\times y_{N2j'} + \eqref{eq:en4}\times x_{N1j}$ and $\eqref{eq:en5}$:
\begin{align}
&(EN_{1j1} - EN_{2j1}) \times y_{N2j'} + (EN_{1j4} - EN_{2j4})\times x_{N1j} \nonumber \\
&= s\alpha(x_{N1j}y_{N2j'} - x_{N2j}y_{N2j'} + y_{N1j'}x_{N1j} - y_{N2j'}x_{N1j}) \nonumber \\
&= s\alpha(x_{N1j}y_{N1j'} - x_{N2j}y_{N2j'}) \nonumber \\
&= (EN_{1j5} - EN_{2j5}) \mod p \label{eq:lin-xy-2}~.
\end{align}

Compare $\eqref{eq:en2}\times x_{N1j'} + \eqref{eq:en3}\times y_{N2j}$ and $\eqref{eq:en6}$:
\begin{align}
&(EN_{1j2} - EN_{2j2}) \times x_{N1j'} + (EN_{1j3} - EN_{2j3})\times y_{N2j} \nonumber \\
&= s\alpha(y_{N1j}x_{N1j'} - y_{N2j}x_{N1j'} + x_{N1j'}y_{N2j} - x_{N2j'}y_{N2j}) \nonumber \\
&= s\alpha(x_{N1j'}y_{N1j} - x_{N2j'}y_{N2j}) \nonumber \\
&= (EN_{1j6} - EN_{2j6}) \mod p \label{eq:lin-xy-3}~.
\end{align}

Compare $\eqref{eq:en2}\times x_{N2j'} + \eqref{eq:en3}\times y_{N1j}$ and $\eqref{eq:en6}$:
\begin{align}
&(EN_{1j2} - EN_{2j2}) \times x_{N2j'} + (EN_{1j3} - EN_{2j3})\times y_{N1j} \nonumber \\
&= s\alpha(y_{N1j}x_{N2j'} - y_{N2j}x_{N2j'} + x_{N1j'}y_{N1j} - x_{N2j'}y_{N1j}) \nonumber \\
&= s\alpha(x_{N1j'}y_{N1j} - x_{N2j'}y_{N2j}) \nonumber \\
&= (EN_{1j6} - EN_{2j6}) \mod p \label{eq:lin-xy-4}~.
\end{align}

Compare \eqref{eq:en1} and \eqref{eq:en2}:
\begin{align}
&(EN_{1j1}-EN_{2j1}) \times s\alpha(y_{N1j}-y_{N2j}) \nonumber \\
=\ &(EN_{1j2}-EN_{2j2}) \times s\alpha(x_{N1j}-x_{N2j}) \mod p~, \nonumber \\
\Rightarrow \ &(EN_{1j2}-EN_{2j2})(x_{N1j}-x_{N2j}) \nonumber \\
-\ &(EN_{1j1}-EN_{2j1})(y_{N1j}-y_{N2j}) = 0 \mod p \label{eq:lin-xy-5}~.
\end{align}

Similarly, compare \eqref{eq:en2} and \eqref{eq:en3}, and \eqref{eq:en3} and \eqref{eq:en4}:
\begin{align}
&(EN_{1j3}-EN_{2j3})(y_{N1j}-y_{N2j}) \nonumber \\
-\ &(EN_{1j2}-EN_{2j2})(x_{N1j'}-x_{N2j'}) = 0 \mod p~, \label{eq:lin-xy-6} \\
&(EN_{1j4}-EN_{2j4})(x_{N1j'}-x_{N2j'}) \nonumber \\
-\ &(EN_{1j3}-EN_{2j3})(y_{N1j'}-y_{N2j'}) = 0 \mod p \label{eq:lin-xy-7}~.
\end{align}

Consider Equations \eqref{eq:lin-xy-1}---\eqref{eq:lin-xy-7} for all $j=1,\dots, 4;\enspace j'=(j\mod 4)+1$. There are $28$ linear (modular) equations in the $16$ unknowns $(x_{N1j},y_{N1j}),\allowbreak (x_{N2j},y_{N2j}); \allowbreak \enspace j=1,\dots,4$. This can be treated as a linear system of equations with elements from the field $\mathbb{Z}_p$, and standard techniques from linear algebra such as Gaussian Elimination can be applied to find solutions for $X$ in $\mathbb{Z}_p$.

\subsubsection{Existence of a unique solution} 

Suppose we represent Equations \eqref{eq:lin-xy-1}---\eqref{eq:lin-xy-7} using matrix notation as $PX=Q$, where $dim(P)=28\times 16$, $dim(X)=16\times 1$, $dim(Q)=28\times 1$, and vector $X$ represents the 16 unknown quadrant vertices of $N_1,N_2$. We observed that $rank(P)\leq 13<16$, and the RV cannot obtain unique solutions for $X$ from this system. 

Hence we propose a modification to our attack such that $rank(P)$ equals the number of unknowns. Previously, considering only $N_1,N_2$ gave us 28 equations and $8\times2=16$ unknowns. If we instead consider $N_1,N_2,N_3$ and take ${3\choose2} = 3$ pairwise combinations, we end up with $28\times3=84$ equations and $8\times3=24$ unknowns (which is slightly better). But we observed that in some cases, the resulting $84\times24$ matrix $P$ had rank $23<24$. Next, considering $N_1,N_2,N_3,N_4$ and taking ${4\choose2}=6$ pairwise combinations gives us $28\times6 = 168$ equations and $8\times4=32$ unknowns. We observed (from experiments described in Section \ref{sec:experimental-results}) that the corresponding $168\times32$ matrix $P$ always had rank $32$, and an RV can therefore solve this system to get the unique values (in $\mathbb{Z}_p$) of quadrant vertices for $N_1,\dots,N_4$. One can proceed further and consider more $N_i$, but that would be redundant since rank already equals the number of unknowns.

We now formalize the above idea. Let the linear system defined by Equations \eqref{eq:lin-xy-1}---\eqref{eq:lin-xy-7} (for vertices of $N_1,N_2$) be denoted by

\begin{equation}
\left[
\begin{array}{c;{2pt/2pt}c}
\\
P_{N1N2} & P_{N2N1} \\
\\
\end{array}
\right]
\left[
\begin{array}{c}
X_{N1} \\ \hdashline[2pt/2pt]
X_{N2}
\end{array}
\right]
=
\left[
\begin{array}{c}
\\
Q_{N1N2} \\
\\
\end{array}
\right].
\label{eq:px=q:pair}
\end{equation}

Here $P_{N1N2},X_{N1}$ and $P_{N2N1},X_{N2}$ are submatrices corresponding to unknown vertices of $N_1$ and $N_2$, respectively. Note that $ dim(P_{N1N2})\allowbreak=dim(P_{N2N1})=28\times8$, $dim(X_{N1})\allowbreak=dim(X_{N2})=8\times1$, $dim(Q_{N1N2})=28\times1 $.
In the same manner, take all ${4\choose2}=6$ pairwise combinations $N_i,N_i';\enspace 1\leq i<i'\leq 4$ from $N_1,N_2, \allowbreak N_3,N_4$ and compute $P_{NiNi'}, \allowbreak X_{Ni}, \allowbreak P_{Ni'Ni}, \allowbreak X_{Ni'}, \allowbreak Q_{NiNi'}$. Define a linear system that considers all the above systems simultaneously.
\begin{equation*}
PX=Q~,
\end{equation*}
\begin{equation}
\left[
\begin{array}{c;{2pt/2pt}c;{2pt/2pt}c;{2pt/2pt}c}
P_{N1N2} & P_{N2N1} & 0 & 0  \\ \hdashline[2pt/2pt]
P_{N1N3} & 0 & P_{N3N1} & 0\\ \hdashline[2pt/2pt]
P_{N1N4} & 0 & 0 & P_{N4N1}\\ \hdashline[2pt/2pt]
0 & P_{N2N3} & P_{N3N2} & 0\\ \hdashline[2pt/2pt]
0 & P_{N2N4} & 0 & P_{N4N2}\\ \hdashline[2pt/2pt]
0 & 0 & P_{N3N4} & P_{N4N3}\\ 
\end{array}
\right]
\left[
\begin{array}{c}
X_{N1} \\ \hdashline[2pt/2pt]
X_{N2} \\ \hdashline[2pt/2pt]
X_{N3} \\ \hdashline[2pt/2pt]
X_{N4}
\end{array}
\right]
=
\left[
\begin{array}{c}
Q_{N1N2} \\ \hdashline[2pt/2pt]
Q_{N1N3} \\ \hdashline[2pt/2pt]
Q_{N1N4} \\ \hdashline[2pt/2pt]
Q_{N2N3} \\ \hdashline[2pt/2pt]
Q_{N2N4} \\ \hdashline[2pt/2pt]
Q_{N3N4}
\end{array}
\right].
\label{eq:px=q}
\end{equation}
Here $0$ denotes the zero matrix of dimension $28\times8$, $dim(P)=168\times32$, $dim(X)=32\times1$, $dim(Q)=168\times1$ and $rank(P)$ is experimentally observed to be 32. The RV can solve this system to obtain unique solutions for $X$ (i.e.~quadrant vertices of $N_1,\dots,N_4$) in $\mathbb{Z}_p$. 

Note that there is no restriction here to use equations for the first four quadrants $N_1,\dots,N_4$. The RV can consider equations corresponding to any four distinct $N_i$ and find their underlying vertices.
The above steps are repeated for other quadrants as well, until all of them are recovered. We summarize the attack in Algorithm \ref{algo:rv-quadtree}.
The same idea also allows an RC to recover the quadtree, when it receives $EN$ from RS.

We remark that this attack is purely algebraic and does not make any assumptions about geometry of the region. The same attack would still work even if quadrants in the spatial division were not restricted to rectangles/squares.

\begin{algorithm}[t]
	\small
	\SetAlgoLined\DontPrintSemicolon
	
	\SetKwFunction{algoOne}{Recover\_Quadtree}
	\SetKwProg{myalgoOne}{Procedure}{ : }{}
	
	\SetKwInOut{Input}{Input}\SetKwInOut{Output}{Output}
	\Input{ Size of quadtree $m$, masked quadtree $EN=(EN_1,\dots,EN_m)$}
	\Output{ Underlying quadrant vertices $N=(N_1,\dots,N_m)$ }
	
	\myalgoOne{\algoOne{$m$, $EN$}} {
		
		\While{\textnormal{\texttt{size}}$(EN) > 0$}{
			Pick four random entries $EN_{a}$, $EN_{b}$, $EN_{c}$, $EN_{d}$ and delete them from $EN$ \\
			\For{\textnormal{each of the $4\choose2$ pairwise combinations $(i,i')$ from $a,b,c,d$}} {
					Obtain a linear system in the unknown vertices of $N_i,N_i'$ using equations similar to \eqref{eq:lin-xy-1}---\eqref{eq:lin-xy-7} \\
					Let the corresponding matrices be $P_{NiNi'},\allowbreak X_{Ni},\allowbreak P_{Ni'Ni},\allowbreak X_{Ni'},\allowbreak Q_{NiNi'}$ similar to \eqref{eq:px=q:pair}\\
				}
			Using the above matrices, define the system $PX=Q$ similar to \eqref{eq:px=q} \\
			Solve this system to obtain quadrant vertices corresponding to $N_a,N_b,N_c,N_d$\\
		}
		\textbf{Output}: $(N_1,\dots,N_m)$
	}
	
	\caption{RV recovers quadtree}
	\label{algo:rv-quadtree}
\end{algorithm}

\textbf{Complexity.}
The linear system of equations represented by $PX=Q$, where $dim(P)=168\times 32, dim(X)=32\times 1, dim(Q)=168\times 1$, and all operations are in the field $\mathbb{Z}_p$, can be solved in time $O((\log{p})^2)=O(k_1^2)$ \cite{gaussTimeComp}. We need to repeatedly solve such a system $\lceil m/4 \rceil$ times to recover vertices of all $m$ quadrants. The total asymptotic complexity of this attack is $O(k_1^2m)$. Our attack is efficient in practice, and Table \ref{tab:rs-quadtree-timing} shows the average time taken to recover quadrant vertices for varying tree sizes and security parameters.

\textbf{Remark.}
The aforementioned attack mainly relies on the fact that TRACE uses the same set of $24$ random values $a_{jh}, j = 1,\dots,4;h=1,\dots,6$ throughout all $EN_i, i=1,\dots,m$ (refer to Step 1 of the TRACE protocol in Section \ref{sec:trace-protocol}). However, upon careful observation, one can see that the correctness of the TRACE protocol would still hold if different set of values of $a_{jh}$ were used for each $EN_i$. That is, sample $24\times m$ independent random values $a_{ijh}, i=1,\dots,m;j=1,\dots,4;h=1,\dots,6$ and mask $EN_{ijh}$ with $a_{ijh}$. In the TRACE protocol, these random values are involved only when computing $B_{ij1}$ and $B_{ij2}$ (Step 3). 
Correctness still holds since these values cancel each other out when computing $B_{ij1} - (B_{ij1} \mod \alpha^2)$. 

Therefore, one can modify the TRACE protocol by using a new random $a_{ijh}$ each time when computing $EN_{ijh}$. This is a countermeasure to prevent RCs and RVs from obtaining the secret quadtree because, $EN_{ijh}$ is masked by fresh randomness each time and no information can be obtained about $(x_{Nij}, y_{Nij})$ given the $EN_i$ values (similar to a one-time pad). This simple observation leads to the following lemma.

\begin{lemma}
The above modification to TRACE provides information-theoretic security against any passive adversary who wishes to obtain additional information about the quadtree maintained by RS.
\end{lemma}

However, as we shall see in Section \ref{sec:attack-rs-location}, this modification does not prevent the RS in obtaining locations of RCs and RVs.
We will also later see that a similar countermeasure does not exist for the latter attack. Trying to use fresh randomness there will violate the correctness of the protocol.

\subsection{RS obtains locations of RCs, RVs}
\label{sec:attack-rs-location}

\textbf{RS finds location of RVs.} In Step 3, RS receives $A_i=A_{i\pi(1)} \allowbreak \|A_{i\pi(2)} \allowbreak \| \allowbreak  A_{i\pi(3)} \allowbreak  \| A_{i\pi(4)}, \enspace i=1,\dots,m$, from each RV, that contains masked information about $(x_V,y_V)$. RS knows $EN_{ij}$ but does not know $r_{ij}$ and the random permutation $\pi$ used on the four $A_{ij}$ values. Since there can only be 24 possible choices for $\pi$, the RS can enumerate all of them to try and find $\pi$. 

For each $i$, RS initializes an empty set $S_i$. For each choice of permutation $\rho$ (among the set of all permutations on four elements), RS permutes the four components of $A_i$ according to $\rho$. That is, RS computes 
\begin{align}
A'_i&=A_{i\rho(\pi(1))}\|\allowbreak A_{i\rho(\pi(2))}\|\allowbreak A_{i\rho(\pi(3))}\allowbreak \| A_{i\rho(\pi(4))} \nonumber \\
&= A'_{i1}\|A'_{i2}\allowbreak \|A'_{i3}\|A'_{i4}.
\label{eq:ai-permute-rho}
\end{align}
$A'_i$ corresponds to the original value $A_{i1}\|A_{i2}\allowbreak \|A_{i3}\|A_{i4}$ computed by RV only when $\rho=\pi^{-1}$. To see if the current choice $\rho$ equals $\pi^{-1}$, RS can do the following: eliminate $r_{ij}, \alpha$ from $A'_{ij1}$ and $A'_{ij2}$ to get a linear equation in the unknowns $x_V,y_V$:
\begin{align}
&A'_{ij1}(x_V\cdot EN_{ij2} + y_V\cdot EN_{ij3} + EN_{ij5}) \nonumber \\ 
=\ &A'_{ij2}(x_V\cdot EN_{ij4} + y_V\cdot EN_{ij1} + EN_{ij6}) \mod p~. 
\label{eq:aij-rs}
\end{align}
In this way, RS can obtain four linear equations for $j=1,\dots, 4$, in the two unknowns $x_V,y_V$. Two of these equations can be used to solve and find $x_V,y_V$ (if a solution does not exist, move to the next choice of $\rho$). The remaining two equations can be used to check if the values of $x_V,y_V$ previously obtained are consistent. If so, then with high probability RS can infer that $\rho=\pi^{-1}$; add this solution to set $S_i$. If it is not consistent discard it and check the next permutation choice for $\rho$. Note that $S_i$ always contains the original $x_V,y_V$ chosen by RV since this solution satisfies the consistency checks when $\rho=\pi^{-1}$. In rare cases it could be possible that a false positive also passes these consistency checks for a different $\rho$ and is added to $S_i$.

The above procedure is discussed only for one value of $i$. There are $m$ such $A_i$'s received by RS (in general $30\leq m \leq 100$ \cite[Section \uppercase\expandafter{\romannumeral 6}]{tracewang2018}), and the original $x_V,y_V$ is present in each $S_i$. Moreover, it is highly unlikely that the same false positive appears in every $S_i$. Therefore, it is very likely that there is only one common element present in all $S_i$ (this probability increases with $m$), and that would be the required location of $RV$. Once again, there are no assumptions made regarding geometry of the spatial region. We summarize the attack in Algorithm \ref{algo:rs-rv-location}.

\begin{algorithm}[t]
	\small
	\SetAlgoLined\DontPrintSemicolon
	
	\SetKwFunction{algoOne}{Recover\_Location}
	\SetKwProg{myalgoOne}{Procedure}{ : }{}
	
	\SetKwInOut{Input}{Input}\SetKwInOut{Output}{Output}
	\Input{ $(A_1,\dots,A_m)$ representing masked information about an RV's location }
	\Output{ RV's location $(x_V,y_V)$ }
	
	\myalgoOne{\algoOne{$A_1,\dots,A_m$}} {
		
		\For{$i=1,\dots,m$}{
			$S_i=\phi$ \\
			$\mathcal{P} = \textnormal{set of all permutations on 4 elements}$ \\
			\For{$\rho \in \mathcal{P}$}{
				\MyComment{Permute the 4 components of $A_i$ using $\rho$} \\
				Compute $A_i'$ according to \eqref{eq:ai-permute-rho} \\
				Obtain a linear system in the unknowns $x_V,y_V$ by substituting $j=1,2$ in \eqref{eq:aij-rs}  \\
				\If{this system does not have a unique solution}{\Continue} 
				\MyComment{Solve this system and check if the unique solution $(x,y)$ also satisfies the two equations obtained when substituting $j=3,4$ in \eqref{eq:aij-rs}} \\
				\If{$(x,y)$ satisfies the consistency check}{$S_i = S_i \cup \{(x,y)\}$}
				\MyComment{Note that $\forall i, (x_V,y_V)\in S_i$}
			}
		\MyComment{With high probability, we have $|\cap_{i=1}^{m}S_i|=1$} \\
		$\{(x_V,y_V)\}=\cap_{i=1}^{m}S_i$\\
		}

		\textbf{Output}: $(x_V,y_V)$
	}	
	\caption{RS recovers location of an RV}
	\label{algo:rs-rv-location}
\end{algorithm}

\textbf{Complexity.}
For each $i=1,\dots,m$, RS enumerates over all 24 possible permutations. In each choice of permutation, RS solves a system of equations in two variables (with all elements being in $\mathbb{Z}_p$) and checks for consistency with two other equations to finally obtain the set $S_i$. The size of each $S_i$ is at most 24. RS later computes the intersection of all $S_i$ to determine the RV's coordinates. All these operations can be done in time $O(k_1^2m)$. Table \ref{tab:rs-rv-timing} shows the average time taken to recover $(x_V,y_V)$ for varying tree size and security parameters.

\textbf{RS finds location of RCs.} In Step 5, RS receives $C_1\|\allowbreak C_2\|\allowbreak C_3\|\allowbreak C_4$ from RC. Recall that RC chooses a square $S_{RC}$ of side $2R$ with its pick-up location at the center. Each $C_j$ corresponds to one vertex of that square, masked in a manner similar to how RV masked its location as $A$ (refer to Step 2). Since we just saw an attack where RS can recover the original underlying location when given such a masking, RS can obtain the 4 vertices corresponding to $S_{RC}$. The center of this square directly gives the pick-up location of RC. 

RS can also find the take-off location of RC. In Step 8, RS receives $C_5$ from RC, which is a masking of RC's take-off location using $EN$ similar to what we have seen in Step 2. Using the same attack as for RV, RS can directly recover RCs location from $C_5$. Note that, in practice, the pick-up and the take-off points are quite close.

\textbf{Remark.} In Section \ref{sec:attack-rc-rv-quadtree} we saw that using fresh randomness for each encrypted quadtree term $EN_i$ did not violate correctness of the protocol. If we try to apply the same argument here, then in Step 2 of the TRACE protocol, $A_{ij1}$ and $A_{ij2}$ have to be masked with different (and fresh) randomness, say $r_{ij1}$ and $r_{ij2}$, respectively (currently, they are both masked by the same $r_{ij}$). But in Step 3 of the TRACE protocol, this would mean $B_{ij1}'$ is masked with $r_{ij1}$, and $B_{ij2}'$ is masked with $r_{ij2}$. Hence $B_{ij}=B_{ij1}'-B_{ij2}'$ would not have a common factor $r_{ij}$, and one cannot infer whether the RV lies inside the quadrant $N_i$ just by checking the sign of $B_{ij}$. Therefore, this approach will violate the correctness of the TRACE protocol, and we believe that other countermeasures for this attack are not straightforward to come up with.

%\subsection{RC obtains location of RV}
%\label{sec:attack-rc-location}
%
%\comsv{See if you can give a more gentle introduction.}
%In Step 7, RC gets $I=r_3(x_{SV}^2+y_{SV}^2+E)$ from RV. It computes $J'=r_3(x_{CP}\cdot x_{SV}+y_{CP}\cdot y_{SV})$. The values unknown to RC are $r_3,x_{SV},y_{SV}$. RC already knows its own location $x_{CP},y_{CP}$ and knows $E=x_{CP}^2+y_{CP}^2-R^2$.
%Recall from Steps 3 and 7 that in order to ensure correctness of the protocol, $|r_3|=k_4$ is significantly large compared to the values of location coordinates. 
%
%\comsv{The following paragraph must be connected to Gauss circle problem. Also talk about similar approach being used in ORide attack.}
%The high-level idea is that RC can efficiently extract GCD of $I$ and $J'$. Let $\mathsf{gcd}(I,J')=g$; $r_3$ divides $g$. Since $r_3$ is expected to be a large factor of $g$, RC can try to obtain a small factor $f$ of $g$ and assume $r_3=g/f$. Then $(x_{SV}^2+y_{SV}^2+E) = \frac{I}{g/f}$. Since $x_{SV},y_{SV}$ are required to be integers RC can find all lattice solutions to the equation $x_{SV}^2+y_{SV}^2 = \frac{I}{g/f} - E$. The false positives can be filtered out by checking consistency using $J'=(g/f)(x_{CP}\cdot x_{SV}+y_{CP}\cdot y_{SV})$. With high probability we expect that RC will end up with one solution to RVs location. The time to enumerate lattice solutions would not be significantly high since $x^2_{SV} + y^2_{SV}$ is small.

%% file: expt-results.tex
\section{Experimental Results}
\label{sec:experimental-results}

In this section, we discuss the experimental setup and other implementation aspects of the attacks \footnote{The implementation can be accessed at \href{https://github.com/deepakkavoor/rhs-attack/tree/trace-attack}{https://github.com/deepakkavoor/rhs-attack/tree/trace-attack}.} mentioned in Section \ref{sec:attack-trace}. Our experiments were implemented using SageMath \cite{sagemath} and run on an Intel Core i5-8250U CPU @ 1.60 GHz with 8 GB RAM running Ubuntu 20.04 LTS.

\subsection{Setup}
\label{sec:expt-setup}
The TRACE paper \cite{tracewang2018} states that setting $(k_1,k_2,\allowbreak k_3,k_4)=(512,160,\allowbreak75,75)$ should be sufficient to ensure that Claims \ref{claim:rs-spatial}, \ref{claim:rs-rc-location}, \ref{claim:rs-rv-location} hold. We also initialize these security parameters with the same values. In addition, we demonstrate the robustness of our attack by performing another set of experiments with larger values $(2048,1000 ,400,400)$ satisfying Equation \eqref{eq:security-params}. Note that our attack is clearly independent of the security of encryption schemes/digital signatures used in TRACE.

The implementation of TRACE protocol from \cite{tracewang2018} does not give any reference to the dataset that was used to create spatial divisions. So, we simulate the creation of an arbitrary  quadtree by first choosing an outermost rectangular quadrant, followed by picking a random center and dividing it into four subquadrants. We repeat this for the smaller quadrants until the number of nodes in the tree is $m$. The TRACE implementation in \cite{tracewang2018} varies $m$ between $28$ and $84$; we set $m=50$ and $m=100$ in our experiments. The attack indeed works for any value of $m$ (recall $m\geq4$) and its success probability increases with $m$.

Integer modular arithmetic is used in all computations. Since the sizes of location coordinates are negligible compared to the security parameters $k_1,\dots,k_4$, the vertices of the outermost quadrant are randomly chosen in the range $[0,2^{20}-1]$ (for $(k_1,k_2,k_3,k_4)=(512,160,75,75)$) and in $[0,2^{50}-1]$ (for $(k_1,k_2,k_3,k_4)=(2048,1000, \allowbreak400,400)$). 

\subsection{RCs, RVs recover Quadtree}
\label{sec:expt-rv-quadtree}
RS computes the encrypted quadtree $EN$ and sends it to an RV as described in Section \ref{sec:trace-protocol}. Next, RV carries out the attack described in Section \ref{sec:attack-rc-rv-quadtree}. We perform 20 iterations of this attack, and in each iteration, the RS generates a fresh random quadtree (as described in Section \ref{sec:expt-setup}) and computes $EN$ accordingly. We observed that in \emph{all} iterations, RV was able to recover the exact values of \emph{all} quadrant vertices every time. 
We repeat the same for different choices of $m$ and security parameters, and tabulate the average time taken to recover the quadtree in Table \ref{tab:rs-quadtree-timing}.
Since the same attack allows an RC to recover the quadtree, similar experimental statistics can be expected in this case.

\begin{table}
\renewcommand{\arraystretch}{1.5}
\centering
\begin{tabular}{|c|c|c|}
\hline
Security parameters & \multicolumn{2}{c|}{Size of quadtree $m$} \\
\cline{2-3}
$(k_1,k_2,k_3,k_4)$ & \ \ \ 50 & \ \ \ 100 \\
\hline
$(512, 160, 75, 75)$ & \ \ \ 55.686 \ \ \  & \ \ \ 108.566 \ \ \ \\
\ \ \ $(2048, 1000, 400, 400)$\ \ \  & \ \ \ 2341.836\ \ \  & \ \ \ 4771.549\ \ \ \\
\hline
\end{tabular}
\vspace{0.1in}
\caption{Time taken (in seconds) for an RV to recover quadtree, averaged over 30 iterations.}
\label{tab:rs-quadtree-timing}
\end{table}

\begin{table}
	\renewcommand{\arraystretch}{1.5}
	\centering
	\begin{tabular}{|c|c|c|}
		\hline
		Security parameters & \multicolumn{2}{c|}{Size of quadtree $m$} \\
		\cline{2-3}
		$(k_1,k_2,k_3,k_4)$ & \ \ \ 50\ \ \  & \ \ \ 100 \ \ \ \\
		\hline
		$(512, 160, 75, 75)$ & \ \ \ 0.206\ \ \  & \ \ \ 0.402 \ \ \ \\
		\ \ \ $(2048, 1000, 400, 400)$\ \ \  & \ \ \ 7.461\ \ \  & \ \ \ 14.778\ \ \ \\
		\hline
	\end{tabular}
\vspace{0.1in}
	\caption{Time taken (in seconds) for RS to recover an RV's location, averaged over 30 iterations.}
	\label{tab:rs-rv-timing}
\end{table}

\subsection{RS recovers locations of RCs and RVs}
\label{sec:expt-rs-rv}
The location $(x_V,y_V)$ of an RV is randomly chosen within the outermost quadrant. We simulate the exchange of messages between RS and this particular RV, following the steps of TRACE protocol (Section \ref{sec:trace-protocol}). Next, RS carries out the attack described in Section \ref{sec:attack-rs-location}. We perform 30 iterations of the attack with freshly generated (random) values for quadtree and $(x_V,y_V)$ in each iteration. We observed that in \emph{all} iterations, RS was able to recover the exact location of the RV. That is, $|\cap_{i=1}^{m}S_i|$ was exactly 1, and the recovered coordinate was same as the RV's location in all iterations (refer Algorithm \ref{algo:rs-rv-location}).
We repeat the same for different choices of $m$ and security parameters, and tabulate the average time taken to recover RV's location in Table \ref{tab:rs-rv-timing}.

The attack to recover an RC's location is exactly the same as that for an RV. Since we assume the distribution of RC's location to be random as well, the same statistics also hold true when RS recovers the location of an RC.

%% file: related_work.tex
\section{Related Work}
\label{sec:related-work}

We briefly mention the prior works in privacy-preserving ride-hailing services. Since these works use fundamentally different ideas (such as homomorphic encryption, garbled circuits) compared to TRACE (which relies on random masking), our attack does not directly apply to these works. %This excludes ride-\textit{sharing} services, in which a driver has to dynamically accommodate multiple riders from different pick-up locations in a single ride.

{\em  PrivateRide} by Pham et al.~\cite{Pham2017PrivateRideAP} is the first work that provides a practical solution towards privacy in ride-hailing systems. The locations and identities of riders are hidden using cloaked regions and anonymous credentials. They use efficient cryptographic primitives to ensure privacy of sensitive information.  
{\em ORide} by Pham et al.~\cite{ORidePaper} offers accountability guarantees and secure payments along with privacy of riders and drivers. They use homomorphic encryption to compute the Euclidean distance and identify the closest driver in a zone. \cite{orideAttackDK} proposes a modification to ORide to ensure location privacy of responding drivers in the region with respect to a rider.

Zhao et al.~\cite{Zhao2019GeolocatingDA} conduct a study on leakage of sensitive data in ride-hailing services. They analyze APIs in non-privacy preserving apps provided to drivers by Uber and Lyft.

{\em pRide} by Luo et al.~\cite{pRideLuo} proposes a privacy-preserving solution involving two non-colluding servers, one of them being the RS and the other a third-party Crypto Provider (CP). They use road network embedding in a higher dimension to approximate shortest distance over road networks. The homomorphically computed (approximate) distances are compared using a garbled circuit.  Their scheme provides higher ride-matching accuracy than \textit{ORide} while being computationally efficient.
{\em lpRide} by Yu et al.~\cite{lpRideYu} improves upon {\em pRide} by eliminating the need for a second Crypto Provider.  They use a modified version of Paillier cryptosystem for encrypting locations of riders and drivers. However, \cite{lpRideAttackSV} proposed an attack on the modified Paillier scheme used in lpRide, allowing the service provider to recover locations of all riders and drivers in the region.

\textit{EPRide} by Yu et al.~\cite{epRideYu} uses an efficient approach to compute the exact shortest road distance using road network hypercube embedding. They use somewhat homomorphic encryption over packet ciphertexts to achieve high ride-matching accuracy and efficiency, reporting significant improvements over \textit{ORide} and \textit{pRide}. Xie et.~al.~\cite{xieTifs2021} improve upon \textit{pRide} by combining the idea of road network embedding with cryptographic constructs such as Property-preserving Hash. They eliminate the need for a trusted third-party server to compute shortest distances.

Lu et al. \cite{LuPPSP} proposed a protocol for Privacy-Preserving Scalar Product (PP-SP) in 2013, which allows two parties $P_0$ and $P_1$ (having input vectors $\overrightarrow{a}$ and $\overrightarrow{b}$, respectively) to jointly compute the scalar product $\overrightarrow{a} \cdot \overrightarrow{b}$ such that no information about $P_i$'s input is revealed to $P_{1-i}$ (other than what is revealed by the output itself), for $i\in \{0,1\}$. Their protocol was claimed to achieve information-theoretic security using random masking, and does not make use of any computational assumptions.
However, in 2019, \cite{Schneider19}
proposed an attack on the PP-SP protocol of Lu et al. and showed that it is impossible to construct a PP-SP protocol without the use of computational hardness assumptions. These attacks are based on constructing distinguishers that leak additional information about the other party's secrets than what the output should reveal. While the TRACE protocol is motivated by the designs of the PP-SP protocols of Lu et al., we would like to stress that the application context, i.e., privacy-preserving ride-hailing services, is different in our setting, and hence the privacy requirements differ too. The main goal of our attacks on the TRACE protocol is the complete recovery of secret locations rather than just distinguishing them from uniform random values, and hence the attack techniques are also different. Note that the anonymity of users' locations is the main requirement for a PP-RHS, and not just indistinguishability from uniform random values. Hence, the attack in \cite{Schneider19} does not necessarily imply our results, though it certainly provides the motivation for a deeper investigation such as our work. 

Also, the impossibility result of \cite{Schneider19} does not necessarily imply that a PP-RHS cannot be constructed without computational hardness assumptions. For instance, in Section \ref{sec:attack-rc-rv-quadtree}, we showed that our modification to the TRACE protocol, where fresh random values are used for each invocation, prevents RCs and RVs from obtaining the secret quadtree (this is based on an information-theoretic argument similar to that of a one-time pad). Hence, impossibility results for the PP-SP setting do not necessarily translate to the PP-RHS setting.

%% file: conclusion.tex
\section{Conclusion and Future Work}
\label{sec:conclusion-future-work}
In this work we proposed an attack on the privacy-preserving ride-hailing service TRACE. We disproved several privacy claims about TRACE in an honest-but-curious setting. We showed how riders (RCs) and drivers (RVs) can recover the secret spatial division information maintained by the ride-hailing server (RS). We also showed how the RS can recover the exact locations of RCs and RVs. We implemented our attack and evaluated the success probability for different security parameters.
In the future, it would be interesting to propose a modified protocol for TRACE in which all the aforementioned privacy claims hold.
\newline

\noindent
\textbf{Acknowledgements.}
This work was partially funded by the Infosys Foundation Career Development Chair Professorship grant for Srinivas Vivek.